\documentstyle[preprint,aps,epsfig]{revtex}

\def\be{\begin{equation}}
\def\ee{\end{equation}}
\def\ep{\epsilon}
\def\t{\tilde}
\def\sg{\sigma}    
\def\et{\tilde{\epsilon}}

\begin{document}
\draft

\title{Dielectric behavior of oblate spheroidal particles: \\ 
       Application to erythrocytes suspensions} 
\author{J. P. Huang and K. W. Yu}
\address{Department of Physics, The Chinese University of Hong Kong,
 Shatin, NT, Hong Kong}
\maketitle

\begin{abstract}
We have investigated the effect of particle shape on the eletrorotation 
(ER) spectrum of living cells suspensions. 
In particular, we consider coated oblate spheroidal particles and present 
a theoretical study of ER based on the spectral representation theory. 
Analytic expressions for the characteristic frequency as well as the 
dispersion strength can be obtained, thus simplifying the fitting of 
experimental data on oblate spheroidal cells that abound in the 
literature. 
From the theoretical analysis, we find that the cell shape, coating 
as well as material parameters can change the ER spectrum.
We demonstrate good agreement between our theoretical predictions and 
experimental data on human erthrocytes suspensions. 
\end{abstract}
\vskip 5mm 

\newpage

\section{Introduction}

Dielectrophoresis and electrorotation (ER) offer a unique capability of 
monitoring the dielectric properties of dispersions of colloids and 
biological cells. 
Under the action of external fields, these particles exhibit rich 
fluid-dynamic behaviors as well as various dielectric responses. 
It is of importance to investigate their frequency-dependent responses 
to ac electric fields, which yields valuable information on the 
structural (Maxwell-Wagner) polarization effects \cite{Gimsa,Gimsa99}. 
The polarization is characterized by a variety of characteristic 
frequency-dependent changes known as the dielectric dispersion. 
In particular, for the $\beta$-dispersion (also called the Maxwell-Wagner 
dispersion, ranging from KHz to MHz), it may suffice to focus on the 
induced dipole moments of the particles. In this work, we will analyze 
the $\beta$-dispersion based on the spectral representation theory.

In the last two decades, various experimental tools have been developed 
to analyze the polarization of biological cells - dielectric spectroscopy 
\cite{Asami80}, dielectrophoresis \cite{Fuhr} and ER \cite{Gimsa91} 
techniques. Among these techniques, conventional dielectrophoresis and ER 
are usually applied to analyze the frequency dependence of translations 
and rotations of individual cells in an inhomogeneous and rotating electric 
field, respectively \cite{Fuhr,Gimsa91}. Moreover, one is able to monitor 
the cell movements by using automated video analysis \cite{Gasperis} 
as well as light scattering methods \cite{Gimsa99}.
As a matter of fact, the general cause of ER is the existence of a phase 
difference between the field-induced dipole moment and the external 
rotating field, resulting in a desired torque which causes the cells to 
rotate.

In the dilute limit, the ER of individual cell can be predicted by 
ignoring the mutual interaction between the cells. However, the cells may 
aggregate under the influence of the external field. 
In this case the Brownian motion can be neglected, and the cell system 
becomes non-dilute even though it is initially dilute.
As an initial model, we have recently studied the ER of two approaching 
spherical particles in the presence of a rotating electric field 
\cite{Huang02}. We showed that when the two particles approach and 
finally touch, the mutual polarization interaction between the particles 
leads to a change in the dipole moment of individual particles and hence 
the ER spectrum, as compared to that of isolated particles, via the multiple 
image method \cite{Yu00}. 

In this work, we consider further the effects of cell shape on the ER 
spectrum.
In fact, a cell may readily deviate from a perfect spherical shape 
due to many reasons, e.g., under a hydrostatic pressure \cite{Asami99}. 
In addition to a prolate spheroidal shape \cite{JPCM}, there exist cells 
of oblate spheroidal shape, such as human erythrocytes. 

Regarding the spectral representation approach \cite{Bergman}, 
it is a rigorous mathematical formalism of the effective dielectric 
constant of a two-phase composite material. 
Also, the approach was extended to deal with a three-phase material 
\cite{Yuen}.
The spectral approach offers the advantage of the separation of material 
parameters (namely the dielectric constant and conductivity) from the 
cell structure information, thus simplifying the study. From this approach, 
one can readily derive the dielectric dispersion spectrum, 
with the dispersion strength as well as the characteristic frequency 
being explicitly expressed in terms of the structure parameters and the 
materials parameters of the cell suspension \cite{Lei}.
In the present work, the analytic expression for the characteristic 
frequency is derived with the aid of the spectral representation 
approach, thus simplifying the fitting of experimental data. 
We show good agreement with experimental data on human erythrocytes 
\cite{Gimsa94}.

\section{FORMALISM}

Consider a suspension of spheroidal particles of complex dielectric constant 
$\t{\ep}_1$ coated with a shell of $\t{\ep}_s$ dispersed in a host 
medium of $\t{\ep}_2$, where
 \be
 \t{\ep}=\ep+\sigma/i2\pi f\nonumber
 \ee 
with $f$ being the frequency of the applied field ${\bf E}_0=E_0 \hat{\bf 
z}$. 
The depolarization factors of the spheroidal particles is described by 
a sum rule
 \be
 L_z+2L_{xy}=1\nonumber
 \ee 
where $L_z (1/3<L_z<1)$ and $L_{xy}$($=L_x=L_y$) are the depolarization 
factors along the $z$- and $x$- (or $y$-) axes of the oblate spheroidal 
particle, respectively. $L_z=L_{xy}=1/3$ just indicates sphere, 
while $L_z=1$ disk.

The phenomenon of ER is based on the interaction between a rotating 
electric field ${\bf E}$ and the induced dipole moment ${\bf M}$. 
The dipole moment of the particle arises from the induced charges that 
accumulate at the interface of the particle. 
In circularly polarized fields, the axis of lowest polarizability, 
namely $z-$axis of the present particle, should be oriented 
perpendicular to the plane of field rotation \cite{Gimsa2001}. 

The angle between ${\bf M}$ and ${\bf E}$ is denoted by $\theta$, 
where $\theta=\omega \times time$ and $\omega=2\pi f$ is angular velocity 
of the rotating electric field. 
The torque acting on the particle is given by the vector cross product 
between the electric field and the dipole moment, so that only the 
imaginary part of the dipole moment contributes to the ER response.
In the steady state, the frequency-dependent rotation speed $\Omega(f)$, 
which results from the balance between the torque and the viscous drag, 
is given by
   \begin{eqnarray}
\Omega(f) &=& -F(\ep_2, E,\eta)Im[\t{b}_x \langle\cos^2\theta \rangle
 + \t{b}_{y} \langle\sin^2 \theta\rangle], \nonumber \\
  &=& -F(\ep_2, E,\eta)Im[\t{b}_{x}]=-F(\ep_2, E,\eta)Im[\t{b}_{y}]\equiv 
-F(\ep_2, E,\eta)Im[\t{b}_{xy}], 
   \end{eqnarray}
where $Im[\cdots]$ indicates taking the imaginary parts of $[\cdots]$,  
the angular brackets denote a time average, and $F$ is a coefficient 
which is proportional to the square of magnitude of field $E^2$, 
but being inversely proportional to the dynamic viscosity $\eta$ 
of the host medium. For spherical cells, $F=\ep_2 E^2/2\eta$. 
Regarding $\eta$, spin friction expression suffices for spherical 
particle. 
However, for spheroidal particles or many particles interacting in a 
suspension, we must consider the more complicated suspension 
hydrodynamics. 
Since the angular velocity of the rotating field is much greater than the 
electrorotation angular velocity, i.e., $\omega \gg \Omega$, the time 
averages are just equal to 1/2. For a single coated spheroidal particle, 
the dipole factor $\t{b}_z$ is given by \cite{JPCM,Gao}
 \be
\t{b}_{xy}=\frac{1}{3}\frac{(\t{\ep}_s-\t{\ep}_2)
 [\et_s+L_{xy}(\et_1-\et_s)]+(\et_1-\et_s)y[\et_s+L_{xy}(\et_2-\et_s)]}
 {(\et_s-\et_1)(\et_2-\et_s)yL_{xy}(1-L_{xy})+[\et_s+(\et_1-\et_s)L_{xy}]
 [\et_2+(\et_s-\et_2)L_{xy}]}
 \ee 
where $y$ is the volume ratio of the core to the whole coated spheroid.
  
We are now in a position to represent $b_{xy}$ in the spectral 
representation. In what follows, we will show that from the spectral 
representation, we can obtain the analytic expression for the 
characteristic frequency at which the maximum ER velocity occurs. 
Let $\et_1=\et_2(1-1/\t{s})$, and assume
$x=\et_s/\et_2$, we obtain
   \be
   \t{b}_{xy}=NP+\frac{F_1}{\t{s}-s_1}
   \label{bxy}
   \ee 
where $NP$ denotes the nonresonant part \cite{Yuen} which vanishes 
in the limit of unshelled spheroidal cells. 
In Eq.(\ref{bxy}), the various quantities are given by
 \begin{eqnarray}
s_1&=&\frac{L_{xy}\{1-(1-x)[1-(1-L_{xy})(1-y)]\}}{x+(1-L_{xy})
 L_{xy}(1-x)^2(1-y)},\nonumber\\
F_1&=&\frac{-x^2y}{3[x+(1-L_{xy})L_{xy}(1-x)^2(1-y)]^2},\nonumber\\
NP&=&\frac{[L_{xy}(1-x)+x](1-x)(1-y)}{3[-x-(1-L_{xy})
 L_{xy}(1-x)^2(1-y)]}.\nonumber
\end{eqnarray}
 
Note that we have assumed $x$ to be a real number, that is, 
$x\approx \sigma_s/\sigma_2$, which will be justified below. 
After substituting $\et=\ep+\sg/i2\pi f$ into Eq.(\ref{bxy}), 
we rewrite $\t{b}_{xy}$ after simple manipulations
 \be
 \t{b}_{xy}=(NP+\frac{F_1}{s-s_1})+\frac{\delta\ep}{1+if/f_{c}}
 \ee
with $s=(1-\ep_1/\ep_2)^{-1}$ and $t=(1-\sg_1/\sg_2)^{-1}$, 
where the dispersion magnitude and characteristic frequency admit 
respectively
 \begin{equation}
 \delta\ep=F_1\frac{s-t}{(t-s_1)(s-s_1)},
\label{dep}
\end{equation}
\begin{equation}
f_{c}=\frac{1}{2\pi}\frac{\sg_2}{\ep_2}\frac{s(t-s_1)}{t(s-s_1)}.
\label{fc}
 \end{equation}

\section{Numerical result}

In Fig.1(a)-(c), the dependence of the pole, characteristic frequency 
and dispersion magnitude on the parameter $x$ are investigated for 
different 
particle shape. For clarity, we set $z=1/y$ from now on. It is evident 
that both $s_1$ and $f_c$ decreases monotonically as $x$ increases; 
larger depolarization factor leads to smaller $s_1$ and $f_c$. 
However, for increasing $x$, $\delta \epsilon$ increases concomitantly; 
also, $\delta \epsilon$ increases while $L_z$ increases.

In Fig.1(d)-(f), we investigate the dependence of the pole, 
characteristic 
frequency and dispersion magnitude on $x$ for different shell thickness. 
For $z=1$, namely without shell, $s_1=0.3(=L_{xy})$ always. 
Increasing $x$ leads to decreasing $s_1$ or $f_c$. 
It is evident that thicker shell yields larger pole or $f_c$ as $x<1$, 
but smaller as $x>1$. In the dispersion magnitude plot, 
$\delta \epsilon$ is constant for the uncoated particles. 
In the case of coated particles, increasing $x$ yields increasing 
$\delta \epsilon$; moreover, larger $z$, smaller $\delta \epsilon$.

In Fig.2, the dependence of material parameters (namely, $s$, $t$, 
$\epsilon_2$, and $\sigma_2$) on $f_c$ and $\delta \epsilon$ is 
discussed. 
$f_c$ is weakly dependent on $s$, but strongly dependent on $t$, 
$\epsilon_2$ and $\sigma_2$. Smaller $|t|$ (or $\epsilon_2$), 
larger $f_c$. However, larger $\sigma_2$, larger $f_c$. 
On the other hand, $\delta \epsilon$ is weakly dependent on $s$ and $t$. 
Also, it is evident that $\epsilon_2$ and $\sigma_2$ play no role in 
$\delta \epsilon$, as Eq.(\ref{dep}) actually shows. 
Note all curves in Fig.2(g) and (h) are respectively overlapped. 
In other words, this predicts that $\epsilon_2$ and $\sigma_2$ plays 
no role in the peak value of the angular velocity (see Fig.4).

In Fig.3, the dependence of $L_z$, $z$ and $x$ on $-Im[b_{xy}]$ are 
investigated, respectively. There is always a peak, the location of 
which is just the characteristic frequency $f_c$. As $L_z$ (or $x$) 
increases, $f_c$ becomes red-shift, while the corresponding peak value 
increases as well. However, a larger $z$ yields smaller peak value, 
while $f_c$ is changed weakly.

In Fig.4, we discuss the dependence of $s$, $t$, $\epsilon_2$ and 
$\sigma_2$ on $-Im[b_{xy}]$. The $s$ effect is small so that it can be 
neglected. Smaller $|t|$ leads to larger $f_c$; at the same time, 
increasing peak value appears. Increasing $\epsilon_2$ yields decreasing 
$f_c$, and the related peak value is  unchanged. 
For increasing $\sigma_2$, $f_c$ increases concomitantly, 
but the peak value remains unchanged as well. 

In Fig.5, the experimental data is extracted from an experiment on 
human erythrocytes \cite{Gimsa94}. Obviously, there is a good agreement 
between theory and experiment. During this fitting, we model the cell 
as oblate spheroid ($L_z=0.65$).

\section{Discussion and conclusion}

In the present work, we have discussed the effect of oblate spheroidal 
particle shape on the ER spectrum. 
In reality, there exist many cells in the form of oblate spheroid, 
such as human erythrocytes \cite{Asami99}. However, there exist only 
a few theories to discuss the ER spectrum of such cells 
(e.g., \cite{Gimsa94} and references therein). 
Our theory is advantageous in that the characteristic frequency at which 
the maximum rotational velocity occurs, is derived analytically, 
which simplifies the fitting of experimental data.

We have assumed $x$, namely the ratio of the shell to host dielectric 
constant, to be a real and positive number. In fact, its imaginary part 
is indeed small.
If we had retained the imaginary part of $x$, two peaks would appear, 
and the conductivity-dominated peak would have occurred at substantially 
lower frequency. Therefore, the neglect of the imaginary part of $x$ is 
to drop the lower frequency peak.
Moreover, according to our calculations, there are two (degenerate) 
sub-dominant poles associated with $b_{xy}$ in the spectral 
representation. 
Thus, in the present work, only one capacitance-dominated peak has been 
shown.

In a recent work \cite{JPCM}, we developed simple equations to describe 
the ER of particles in a suspension from the spectral representation, 
and obtained a good fitting on a coated-bead ER assay by using the coated 
spherical model. 
In this work, an extension to oblate spheroidal case has been made. 
Our theory serves as a basis which describes the parameter dependence 
of the polarization and thereby enhances the applicability of various 
cell models for the analysis of the polarization mechanisms. 
In this connection, the shell-spheroidal cell model may readily be 
extended to multi-shell cell model.
However, we have shown that the multi-shell nature of the cell may have 
a minor effect on the ER spectrum \cite{CTP}.

We have considered the isolated cell case, which is a valid assumption
for low concentration of cells. However, for a higher concentration of 
cells, we should consider the mutual interaction between cells. 
When the volume fraction of the suspension becomes large, the particles 
may aggregate in the plane of the rotating applied field, and the mutual 
interactions between the suspended particles can modify the spin 
friction, 
which is a key to determine the angular velocity of ER. 
For ER of two particles, we have successfully applied the spectral 
representation to deal with the dispersion frequency \cite{Huang02}. 
However, the determination of the spin friction is still lacking. 
For two particles, the basic tool is the reflection method \cite{Happel}, 
being analogous to the multiple image method in electrostatics, 
but being only valid for two particles. For more than two particles, 
we need a first-principles method, e.g., the Green's function 
(Oseen tensor) formulation. 
For a dilute suspension, however, one may adopt the effective medium 
theories \cite{Choy} to capture the effective viscosity of a suspension, 
hence modifying further the ER spectrum.

One may argue that normal erythrocytes may deviate from an oblate 
spheroidal shape. For cells of non-conventional shapes, one may use a 
first-principles approach \cite{LJ}, recently developed to deal with 
particles of rod shape. 
As shown in Ref.\cite{LJ}, the derivation from spheroidal model can be small 
when the cells are rotating with their long axes along the applied field.

In summary, we have presented a theoretical study of electrorotation 
based on the spectral representation theory. 
From the theoretical analysis, we find that cell shape as well as 
the coating can change the characteristic frequency.
By adjusting the cell shape, dielectric properties and the thickness 
of the coating, it is possible to obtain good agreement between our 
theoretical predictions and the experimental data on human erythrocytes 
suspensions.

\section*{Acknowledgments}

This work was supported by the Research Grants Council of the Hong Kong 
SAR Government under grant CUHK 4245/01P. 
K.W.Y. acknowledges useful conversation with Professor G. Q. Gu.

\begin{figure}[h]
\caption{FIG.1. The dependences of pole, characteristic frequency and 
dispersion magnitude on $x$ for $s=1.1$, 
$t=-0.005$, $\ep_2=80\ep_0$, $\sg_2=2.9 \times 10^{-5}S/m$. 
(a)$\sim$(c) $z=2$; 
(d)$\sim$(f) $L_z=0.4$.}
\end{figure}

\begin{figure}[h]
\caption{FIG.2. The dependence of characteristic frequency and dispersion 
magnitude on $x$ for $z=6$, $L_z=0.4$. 
(a) $t=-0.005$, $\ep_2=80\ep_0$, $\sg_2=2.9 \times 10^{-5}S/m$;
(b) $s=1.1$, $\ep_2=80\ep_0$, $\sg_2=2.9 \times 10^{-5}S/m$;
(c) $t=-0.005$, $s=1.1$, $\sg_2=2.9 \times 10^{-5}S/m$;
(d) $s=1.1$, $t=-0.005$, $\ep_2=80\ep_0$;
(e) $t=-0.005$, $\ep_2=80\ep_0$, $\sg_2=2.9 \times 10^{-5}S/m$;
(f) $s=1.1$, $\ep_2=80\ep_0$, $\sg_2=2.9 \times 10^{-5}S/m$;
(g) $t=-0.005$, $s=1.1$, $\sg_2=2.9 \times 10^{-5}S/m$;
(h) $t=-0.005$, $s=1.1$, $\ep_2=80\ep_0$.}
\end{figure}

\begin{figure}[h]
\caption{FIG.3. The dependence of $-Im[b_{xy}]$ on frequency for $s=1.1$, 
$t=-0.005$.
(a) $\ep_2=80\ep_0$, $\sg_2=2.9 \times 10^{-5}S/m$, $z=6$, $x=2$;
(b) $\ep_2=80\ep_0$, $\sg_2=2.9 \times 10^{-5}S/m$, $L_z=0.4$, $x=2$;
(c) $\ep_2=80\ep_0$, $\sg_2=2.9 \times 10^{-5}S/m$, $z=6$, $L_z=0.4$.}
\end{figure}

\begin{figure}[h]
\caption{FIG.4. The dependence of $-Im[b_{xy}]$ on frequency for $x=2$, 
$z=6$, $L_z=0.4$; 
(a) $t=-0.005$, $\ep_2=80\ep_0$, $\sg_2=2.9 \times 10^{-5}S/m$;
(b) $s=1.1$, $\ep_2=80\ep_0$, $\sg_2=2.9 \times 10^{-5}S/m$;
(c) $t=-0.005$, $s=1.1$, $\sg_2=2.9 \times 10^{-5}S/m$;
(d) $t=-0.005$, $s=1.1$, $\ep_2=80\ep_0$.}
\end{figure}

\begin{figure}[h]
\caption{FIG.5. A fitting on human erythrocyte data. 
Experiment: $\sigma_2=12.5\times 10^{-3}$. 
Theory: $s=100$, $t=1.00783$, $\ep_2=80\ep_0$, $z=3$, $x=0.008$, 
$L_z=0.65$, $F/E^2=135$.}
\end{figure}

\newpage
\centerline{\epsfig{file=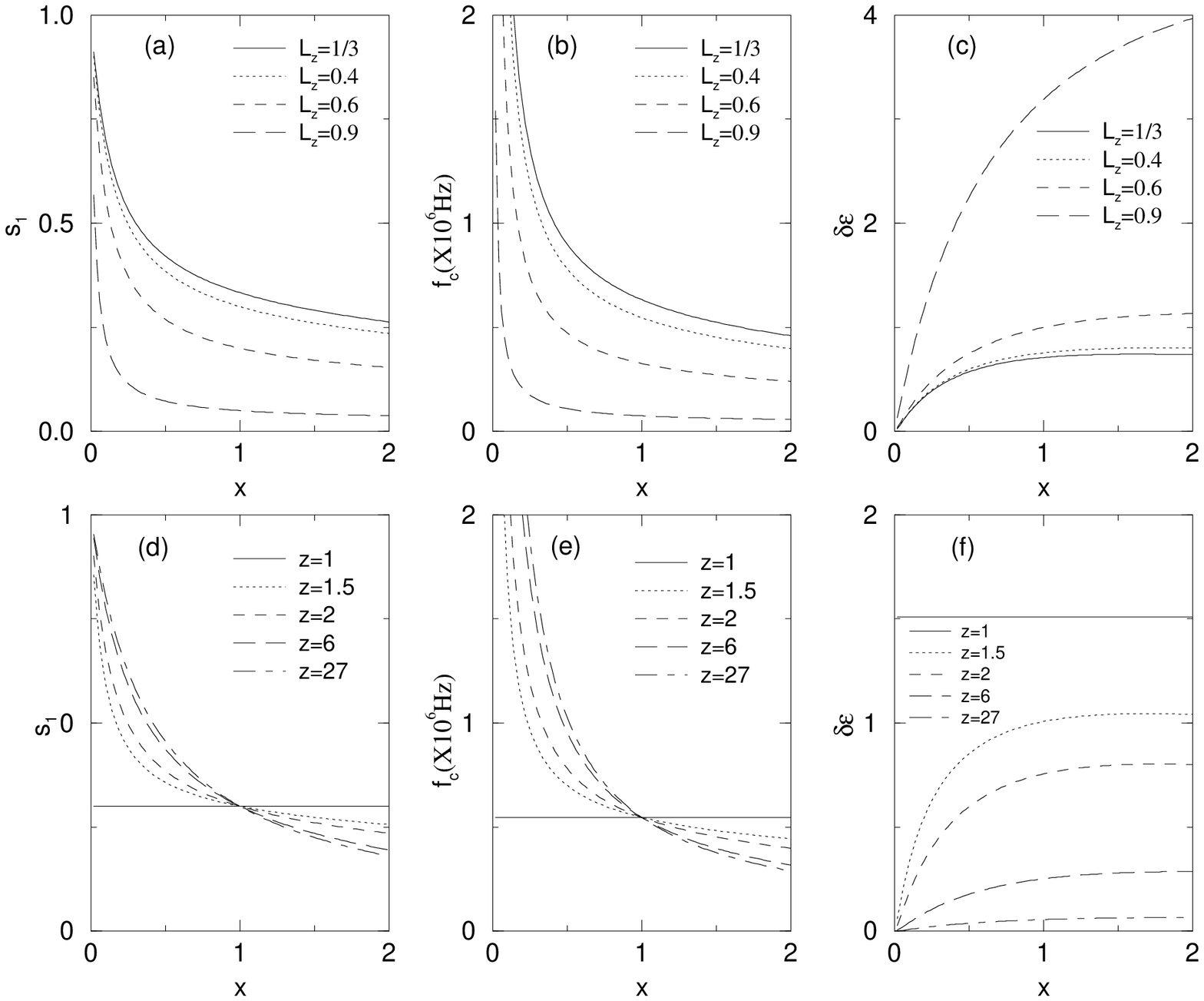,width=\linewidth}}
\centerline{FIG.1}

\newpage
\centerline{\epsfig{file=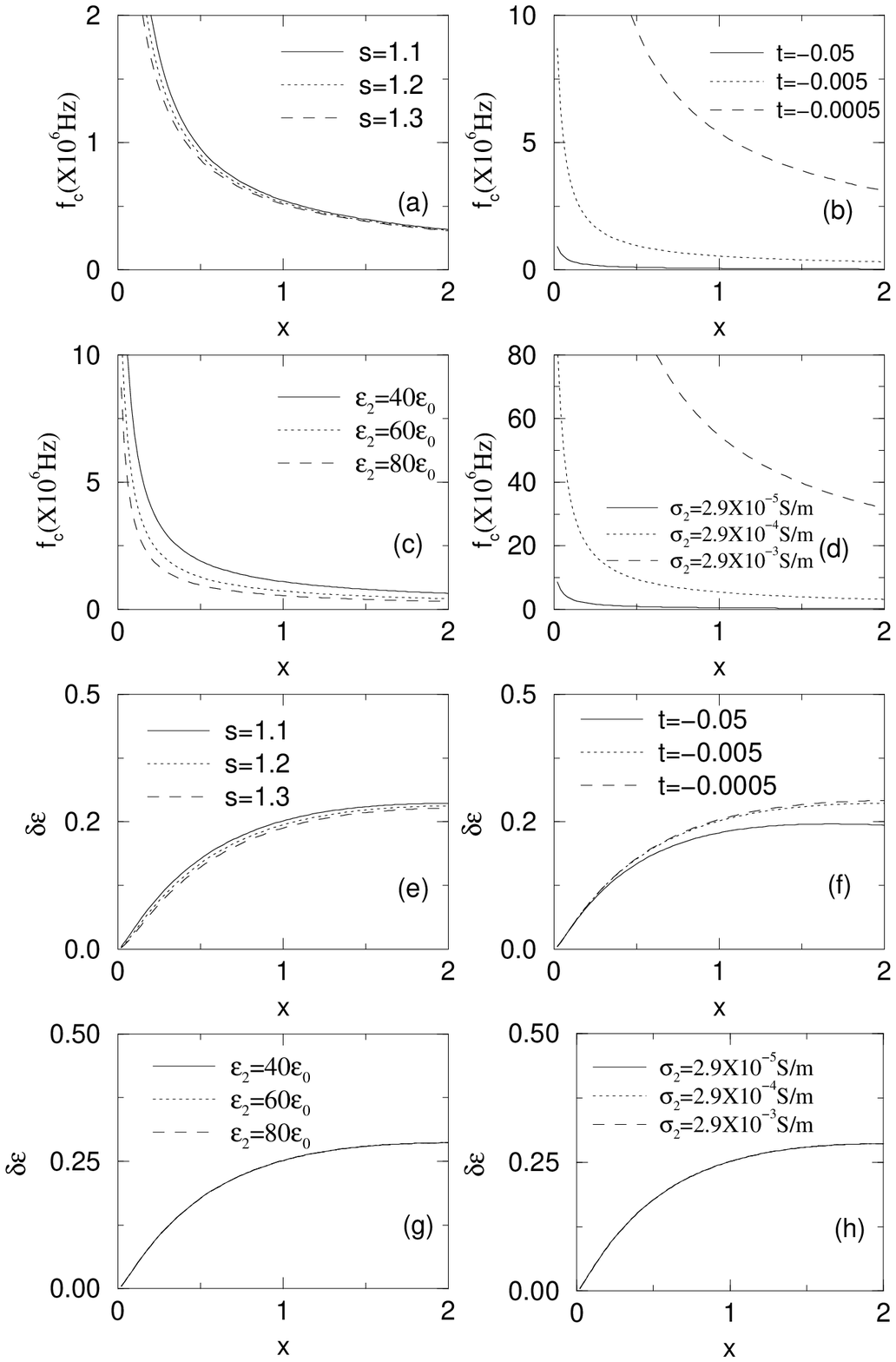,width=\linewidth}}
\centerline{FIG.2}

\newpage
\centerline{\epsfig{file=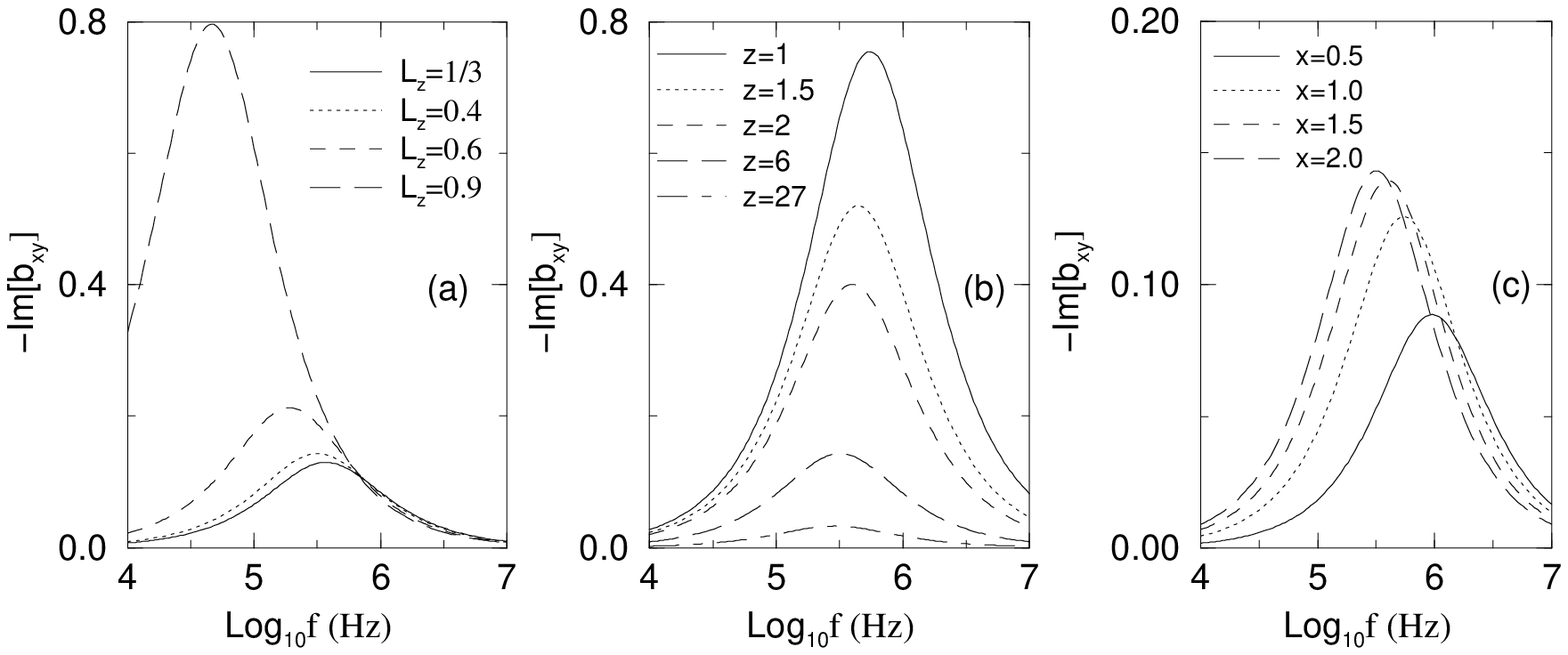,width=\linewidth}}
\centerline{FIG.3}

\newpage
\centerline{\epsfig{file=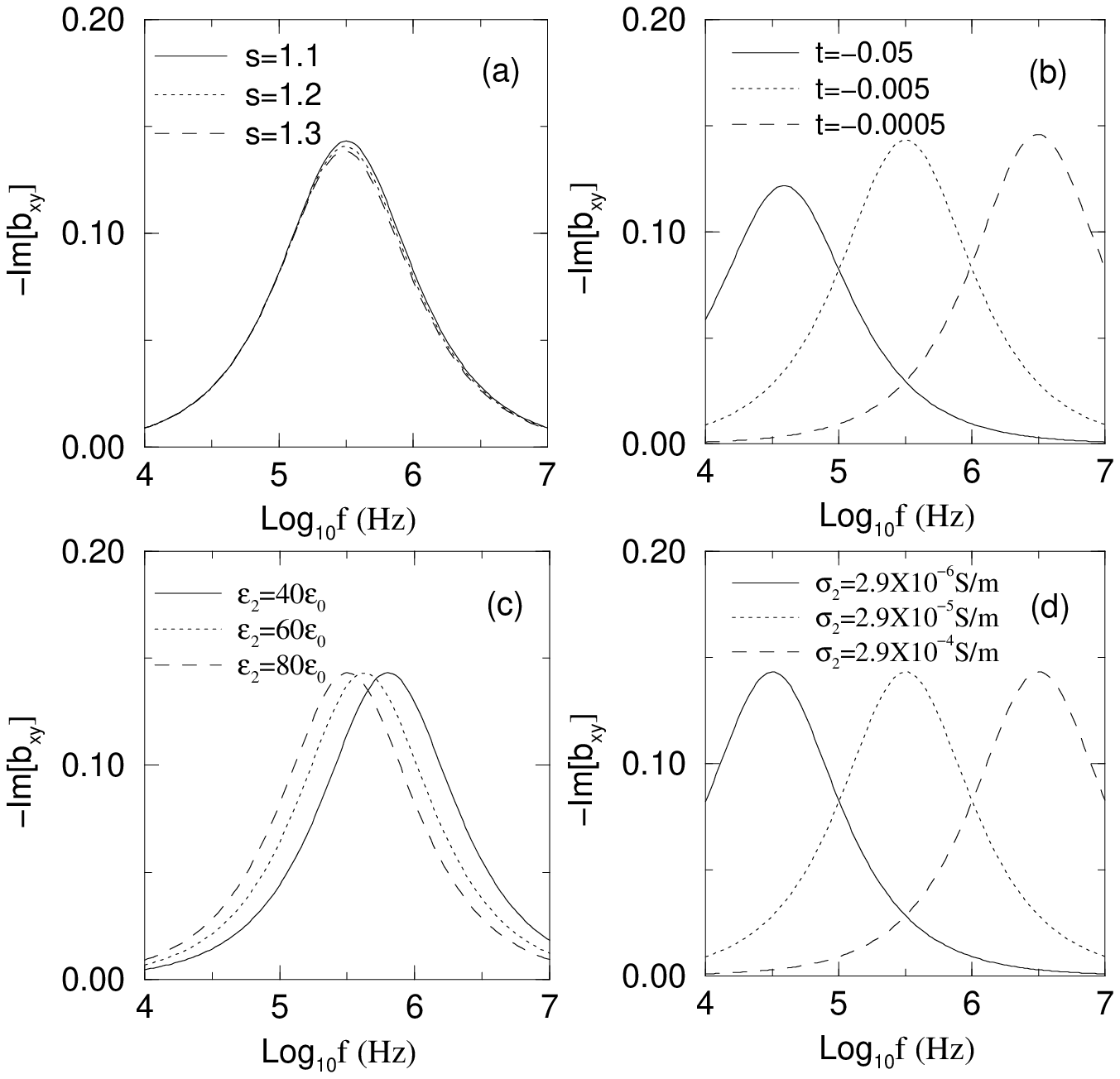,width=\linewidth}}
\centerline{FIG.4}

\newpage
\centerline{\epsfig{file=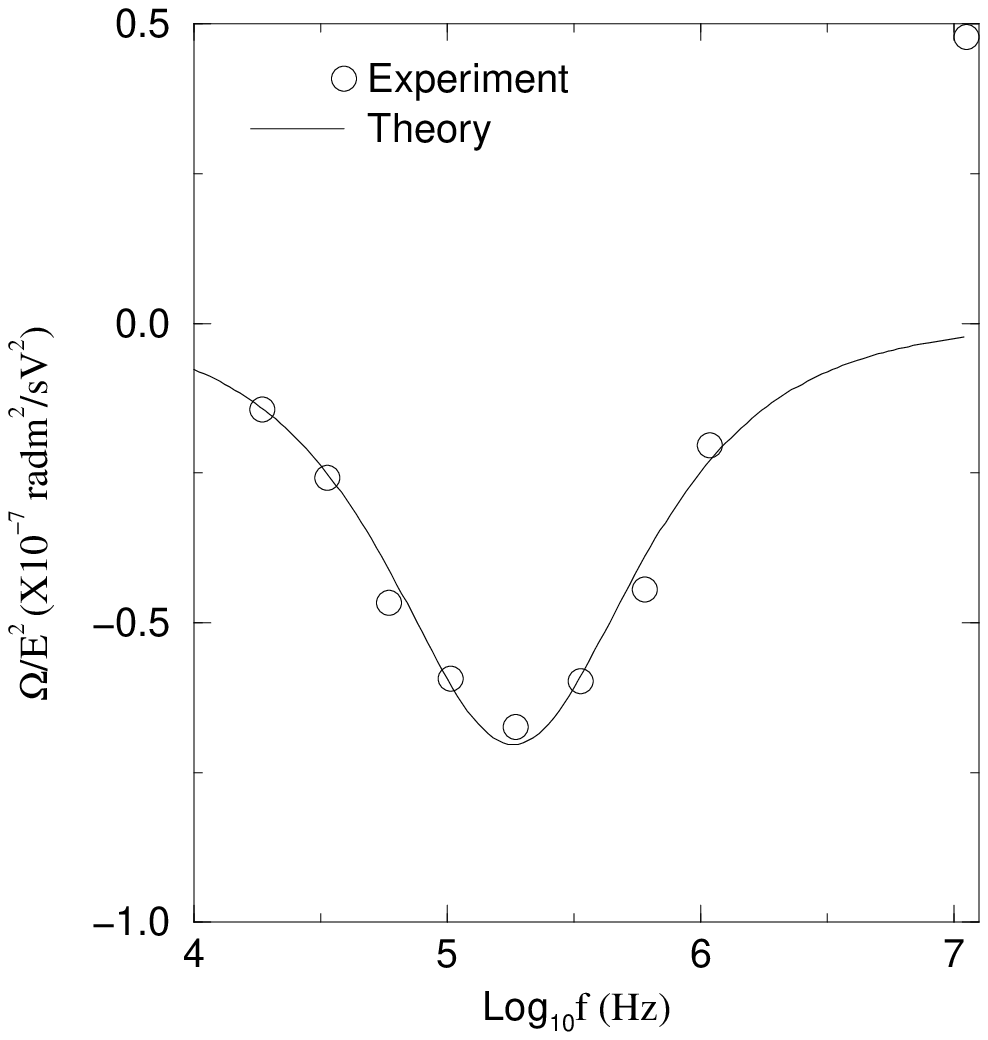,width=\linewidth}}
\centerline{FIG.5}

\end{document}